\documentclass{osa-article}

\journal{oe}

\articletype{Research Article}

\begin{document}

\title{Hard excitation mode of system with optomechanical instability}

\author{Artem Mukhamedyanov,\authormark{1} Alexander A. Zyablovsky,\authormark{1,2,*} and Evgeny S. Andrianov\authormark{1,2}}

\address{\authormark{1}Moscow Institute of Physics and Technology, 9 Institutskiy pereulok, Moscow, Russia, 141700\\
\authormark{2}Dukhov Research Institute of Automatics (VNIIA), 22 Sushchevskaya, Moscow, Russia, 127055\\}

\email{\authormark{*}zyablovskiy@mail.ru} 



\begin{abstract}
System with strong photon-phonon interaction and optomechanical instability are perspective for generation of coherent phonons and photons. Typically, above the threshold of optomechanical instability, the photon intensity increases linearly with pumping. We demonstrate that in such systems, it is possible to achieve hard mode of excitation when jump increase in the photon intensity takes place. We derive the analytical expression determining conditions for such a jump increase. We demonstrate that the hard excitation mode in system with optomechanical instability arises due to an additional phase condition for the existence of a nonzero solution. The discovered hard excitation mode paves the way for creation highly sensitive sensors and optical transistors.
\end{abstract}


In recent years, the creation of sources of coherent phonons — phonon lasers — has attracted much attention \cite{vahala2009phonon,grudinin2010phonon,beardsley2010coherent,jing2014pt,zhang2018phonon,pettit2019optical,chafatinos2020polariton,kabuss2012optically}. These devices are a phonon analogue of conventional (photon) lasers \cite{lamb1999laser}. Like in lasers, the generation of coherent photons and phonons in such devices takes place when the pump rate exceeds the threshold value. After the threshold, the radiation intensity increases much greater with the pump rate than before the threshold. However, the intensity of coherent photons and phonons changes continuously at the generation threshold \cite{vahala2009phonon}. In the conventional laser, such a transition to the laser generation is called as a soft mode of excitation \cite{haken1975generalized,khanin2012principles}.

In ordinary lasers, a jump-like increase of photon intensity at the threshold can take place. Such situation arises in photon lasers with the cavity containing a saturable absorber \cite{khanin2012principles}. In these devices, above the threshold, there are several stable solutions with different intensities of the electromagnetic field in the cavity. For a solution with low intensity, the absorber is unsaturated and the generation does not take place due to the absorber-induced losses. For the solution with high intensity, the absorber is saturated and the generation occurs. Transition from the low- to the high-intensity solution is accompanied by an abrupt change in the intensity of laser radiation. That is called as a hard mode of excitation \cite{khanin2012principles}.

An abrupt change in laser radiation intensity near the threshold can be used to create highly sensitive sensors \cite{melentiev2017plasmonic,ma2014explosives}, optical transistors \cite{zasedatelev2019room,zasedatelev2021single}, ultra-fast optical switch \cite{altug2006ultrafast,englund2008ultrafast,nefedkin2019response}, etc. For example, the sensitivity of sensors operating within the method of intracavity laser spectroscopy is determined by the relative change in laser intensity upon the addition of a single absorbing molecule \cite{melentiev2017plasmonic,ma2014explosives}. The maximum sensitivity is reached near the generation threshold, where the relative change in intensity is maximal \cite{melentiev2017plasmonic,ma2014explosives}. For this reason, lasers, in which the transition to the generation occurs through the hard excitation mode, are useful for the creation of sensors and other applications. However, in optomechanical systems, the hard mode of excitation has not observed.

In this letter, we consider a laser based on an optomechanical system of two optical modes interacting with each other via a phonon mode. We demonstrate, for the first time, that in such a system the hard mode of excitation can take place. We derive the condition for the realization of hard excitation mode and obtain expressions for the generation threshold and the laser curves in the case of both soft and hard modes of excitation. We argue that the hard excitation in the optomechanical system is associated with a phase condition for the existence of a nonzero solution. This mechanism is different from the ones known for the conventional lasers.

We consider a system of two optical modes with frequencies ${\omega _1}$ and ${\omega _2}$ interacting with each other via phonon mode. The frequency of phonon mode is ${\omega _b}$. We use the following optomechanical Hamiltonian \cite{Grudinin} to describe this system

\begin{equation}
\begin{array}{l}
\hat H = \hbar {\omega _{\,1}}\hat a_1^\dag {{\hat a}_1} + \hbar {\omega _{\,2}}\hat a_2^\dag {{\hat a}_2} + \hbar {\omega _{\,b}}{{\hat b}^\dag }\hat b + \\
\hbar g (\hat a_1^\dag {{\hat a}_2}\hat b + {{\hat a}_1}\hat a_2^\dag {{\hat b}^\dag }) + \hbar \Omega ({{\hat a}_1}{e^{i\omega t}} + \hat a_1^\dag {e^{ - i\omega t}})
\end{array}
\label{eq:1}
\end{equation}
Here ${\hat a_{1,2}}$ and $\hat a_{1,2}^\dag$ are the annihilation and creation bosonic operators for the first and the second optical modes, respectively. $\hat b$ and ${\hat b^\dag }$ are the annihilation and creation operators of the phonons that satisfy the bosonic communication relation $\left[ {\hat b,{{\hat b}^\dag }} \right] = 1$. The fourth term in~(\ref{eq:1}) describes the optomechanical interaction between the electromagnetic modes via the phonons \cite{Grudinin}, $g$ is a coupling strength between the modes and the phonons (Frohlich constant). The last term in the Hamiltonian describes the interaction of the first mode with the external EM wave. The intensity of the external wave is determined by $\Omega$. For simplicity, we consider that $g$ and $\Omega$ are positive real quantities.

We use the master equation for density matrix $\hat \rho$ in the Lindblad form \cite{Carmichael,Gardiner} to describe the relaxation processes in the system . Using expressions $\left\langle {\hat A} \right\rangle  = Tr\left( {\hat \rho \hat A} \right)$ and $\left\langle {\frac{{d\hat A}}{{dt}}} \right\rangle  = Tr\left( {\frac{{\partial \hat \rho }}{{\partial t}}\hat A} \right)$ we obtain equations for the average values of the operators ${a_1} = \left\langle {{{\hat a}_1}} \right\rangle$, ${a_2} = \left\langle {{{\hat a}_2}} \right\rangle$ and $b = \left\langle {\hat b} \right\rangle$:
\begin{equation}
\frac{{d{a_1}}}{{dt}} =  - (i{\omega _{\,1}} + {\gamma _1}){a_1} - i g {a_2}b - i \Omega {e^{ - i\omega t}}
\label{eq:2}
\end{equation}

\begin{equation}
\frac{{d{a_2}}}{{dt}} =  - (i{\omega _{\,2}} + {\gamma _2}){a_2} - i g {a_1}{b^*}
\label{eq:3}
\end{equation}

\begin{equation}
\frac{{db}}{{dt}} =  - (i{\omega _{\,b}} + {\gamma _b})b - i g {a_1}{a_2}^*
\label{eq:4}
\end{equation}

To obtain the closed system of differential equations we use the mean-field approximation \cite{Scully} by making substitutions  $\left\langle {{{\hat a}_1}{{\hat b}^\dag }} \right\rangle  \to \left\langle {{{\hat a}_1}} \right\rangle {\left\langle {\hat b} \right\rangle ^*}$, $\left\langle {{{\hat a}_2}\hat b} \right\rangle  \to \left\langle {{{\hat a}_2}} \right\rangle \left\langle {\hat b} \right\rangle$, $\left\langle {{{\hat a}_1}\hat a_2^\dag } \right\rangle  \to \left\langle {{{\hat a}_1}} \right\rangle \left\langle {\hat a_2^\dag } \right\rangle$.

The numerical simulation of the Eqns.~(\ref{eq:2})-(\ref{eq:4}) show that above the lasing threshold, the intensities of the optical modes, $\left| a_{1,2}\right|^2$ and the phonon mode, $\left|b\right|^2$ are constant. Therefore, based on the Equation~(\ref{eq:2}) we conclude that the terms ${a_1}$ and ${a_2 b}$ oscillate with frequency of the external wave, $\omega$. Using this fact, we are looking for a stationary solution of the Eqns.~(\ref{eq:2})-(\ref{eq:4}) in the form ${a_1} = {a_{1st}}{e^{ - i\omega t}}$, ${a_2} = {a_{2st}}{e^{ - i(\omega - \delta\omega)t}}$, ${b} = {b_{st}}{e^{ - i\delta\omega t}}$, where $a_{1st}$, $a_{2st}$, $b_{st}$ denote time independent amplitudes of corresponding variables and $\delta \omega$ is a frequency of generated phonons, which is determined from the equations for stationary solution.

\begin{figure*}[ht]
\centering
\includegraphics[width=\linewidth]{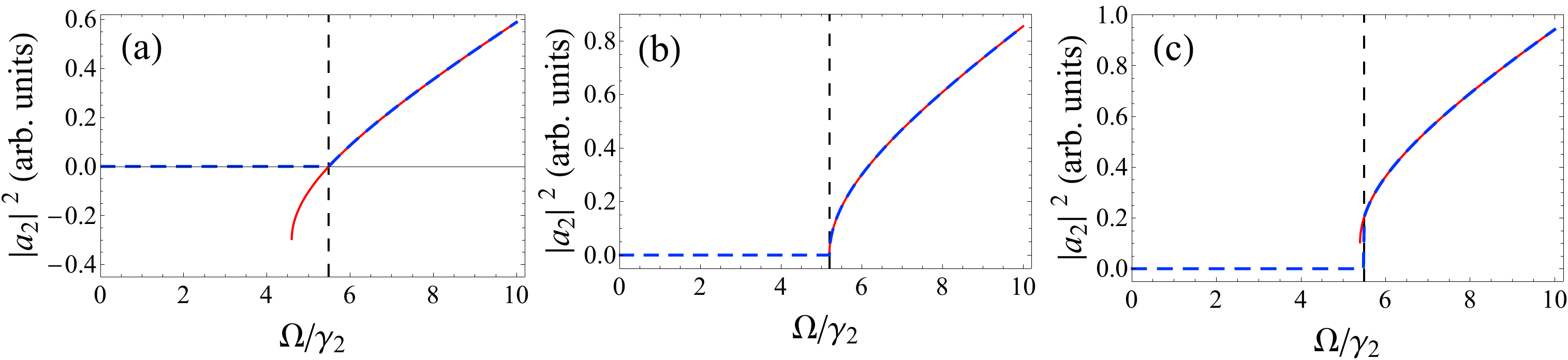}
\caption{The dependence of the intensity of second optical mode, ${\left| {{a_2}} \right|^2}$, on the amplitude of external wave, $\Omega$ (the dashed blue line). The solid red line shows the dependence of ${\left| {{a_2}} \right|^2}$ on $\Omega$ calculated using the Eq.~(\ref{eq:14}). Here $\delta \omega _1 = - 4 \cdot {10^{ - 3}}{\omega _0}$ (a); $\delta \omega _1 = 2 \cdot {10^{ - 3}}{\omega _0}$ (b); $\delta \omega _1 = 4 \cdot {10^{ - 3}}{\omega _0}$ (c). The other parameters are the same for all figures. $\delta \omega _2 = 5 \cdot {10^{ - 3}}{\omega _0}$, $\omega_b = 5 \cdot {10^{-3}}{\omega_0}$, $\gamma_1 = 10^{-2}{\omega_0}$, $\gamma_2 = 10^{-3}{\omega_0}$, $\gamma_b = 10^{-3}{\omega_0}$, $g = \cdot 10^{-2}{\omega_0}$.}
\label{fig:1}
\end{figure*}

The stationary solutions of the Eqns.~(\ref{eq:2})-(\ref{eq:4}) are determined by the following equations:

\begin{equation}
- (i{\delta \omega _{\,1}} + {\gamma _1}){a_{1st}} - i g {a_{2st}}{b_{st}} - i \Omega = 0
\label{eq:5}
\end{equation}

\begin{equation}
- (i{\Delta_{\, 2}} + {\gamma _2}){a_{2st}} - i g {a_{1st}}{b_{st}^*} = 0
\label{eq:6}
\end{equation}

\begin{equation}
- (i{\Delta _{\,b}} + {\gamma _b}){b_{st}} - i g {a_{1st}}{a_{2st}}^* = 0
\label{eq:7}
\end{equation}
where $\delta \omega_{1,2}=\omega_{1,2} - \omega$, $\Delta_2 = \delta \omega_2 + \delta\omega$ and $\Delta_b = \omega_b - \delta\omega$.

One of solutions of the Eqns.~(\ref{eq:5})-(\ref{eq:7}) is given as ${a_{2st}} = {b_{st}}= 0$  and ${a_{1st}} =  - i\,\Omega /\left( {i{\delta \omega _1} + {\gamma _1}} \right)$. This solution (zero solution) corresponds to forced oscillations in the first optical mode under the influence of an external electromagnetic wave. Linear stability analysis of the Eqns.~(\ref{eq:5})-(\ref{eq:7}) shows that the zero solution is stable when

\begin{equation}
{\Omega } < {\Omega _{th}} = \frac{1}{g }\sqrt {\frac{{{\gamma _b}}}{{{\gamma _2}}}} \sqrt {{{\left( {{\delta \omega_1}{\gamma _2} + {\gamma _1}{\Delta _2}} \right)}^2} + {{\left( {{\gamma _1}{\gamma _2} - {\delta \omega _1}{\Delta _2}} \right)}^2}}
\label{eq:7_1}
\end{equation}
When this condition is satisfied, the generation of photons in the second mode and phonons does not occur.

To find a solution describing generation in the second optical mode and phonons, we consider that ${a_{1st,2st}} \ne 0$ and $b_{st} \ne 0$. From Eqns.~(\ref{eq:6}), (\ref{eq:7}) we obtain that

\begin{equation}
{\left| {a_{2st}} \right|^2} = {\left| b_{st} \right|^2}\frac{{{\gamma _b}}}{{{\gamma _2}}}
\label{eq:8}
\end{equation}
and
\begin{equation}
{\Delta _2} = \frac{{\delta {\omega _2} + {\omega _b}}}{{{\gamma _2} + {\gamma _b}}}{\gamma _2}
\label{eq:8_2}
\end{equation}

Then from Eq.~(\ref{eq:6}), we obtain the following expression
\begin{equation}
{a_{1st}} = \frac{{ - \left( {i{\Delta _2} + {\gamma _2}} \right)}}{{i g }}\frac{{{a_{2st}}}}{{{b_{st}^*}}}
\label{eq:9}
\end{equation}

Hereinafter, we use the fact that $\left| {a_2} \right|=\left| {a_{2st}} \right|$ and $\left| b \right|=\left| b_{st} \right|$. We introduce the notation ${a_{2st}}b_{st} = \left| {{a_{2}}} \right|\left| b\right|\exp \left( {i\varphi } \right)$, where $\varphi $ is a total phase of the product of complex amplitudes. Note that since $\Omega$ is real quantity then $\varphi$ determines the phase difference between $a_2 b$ and the external wave. Substituting expression ~(\ref{eq:9}) into Eq.~(\ref{eq:5}) and using the fact that $\frac{{{a_{2st}}}}{{{b_{st}^*}}} = \frac{{\left| {{a_{2}}} \right|}}{{\left| b \right|}}\exp \left( {i\varphi } \right)$, we obtain the following expression

\begin{equation}
\begin{gathered}
  \left( {\frac{{{\gamma _1}{\gamma _2} - {\delta \omega_1}{\Delta _2}}}{g }\frac{{\left| {{a_2}} \right|}}{{\left| b \right|}} + g \left| {{a_2}} \right|\left| b \right| + i\frac{{{\delta \omega _1}{\gamma _2} + {\gamma _1}{\Delta _2}}}{ g }\frac{{\left| {{a_2}} \right|}}{{\left| b \right|}}} \right) \hfill \\
   =  - \Omega \exp \left( { - i\varphi } \right) \hfill \\ 
\end{gathered}
\label{eq:10}
\end{equation}
Dividing Eq.~(\ref{eq:10}) into the real and imaginary parts and using Eq.~(\ref{eq:8}) , we obtain the following expression

\begin{equation}
\Omega \cos \varphi  = \frac{{{\delta \omega_1}{\Delta _2} - {\gamma _1}{\gamma _2}}}{ g } \sqrt {\frac{{{\gamma _b}}}{{{\gamma _2}}}} - g \sqrt {\frac{{{\gamma _2}}}{{{\gamma _b}}}} \left| {{a_2}} \right|^2
\label{eq:11}
\end{equation}

\begin{equation}
\Omega \sin \varphi  = \frac{{{\delta \omega _1}{\gamma _2} + {\gamma _1}{\Delta _2}}}{ g }\sqrt {\frac{{{\gamma _b}}}{{{\gamma _2}}}}
\label{eq:12}
\end{equation}
The Eqns.~(\ref{eq:11}) and (\ref{eq:12}) determine the amplitude of second optical mode, $\left| {{a_2}} \right|$, and the sum phase of the second mode and phonon, $\varphi$.

The nonzero solution exists when $\left| {{a_2}} \right| \geqslant 0$ and $\left| {\sin \varphi } \right| \leqslant 1$. Eq.~(\ref{eq:12}) determines $\varphi$ and plays a role of the phase condition. At the same time, Eq.~(\ref{eq:11}) plays a role of the amplitude condition. The existence of a phase condition leads to the fact that a nonzero solution can appear when, in accordance with Eq.~(\ref{eq:11}), $\left| {{a_2}} \right| \ge const > 0$. In this case, the transition from the zero solution ($a_2 = b = 0$) to the nonzero solution leads to a jump-like increase in the amplitude of the second optical mode. It occurs when the zero solution becomes unstable (see Eq.~(\ref{eq:8})) and corresponds to a hard mode of excitation in the laser. Note that in our case the hard excitation mode is precisely due to the existence of the additional phase condition~(\ref{eq:12}).

To determine the condition for the hard excitation mode, we obtain a solution of Eqns.~(\ref{eq:11}) and (\ref{eq:12}). Using the phase condition~(\ref{eq:12}), we get that the nonzero solution can exist only when
\begin{equation}
\Omega  \geqslant {\Omega _{ex}} = \left| {\frac{{{\delta \omega _1}{\gamma _2} + {\gamma _1}{\Delta _2}}}{ g }} \right|\sqrt {\frac{{{\gamma _b}}}{{{\gamma _2}}}}
\label{eq:13}
\end{equation}
The intensity of the second mode for the nonzero solution is defined as (the solution of Eqns.~(\ref{eq:11} and (\ref{eq:12}))
\begin{equation}
{\left| {{a_2}} \right|^2} = \pm \frac{1}{g }\sqrt {\frac{{{\gamma _b}}}{{{\gamma _2}}}} \sqrt {{{\Omega }^2} -  \Omega _{ex}^2}  + \frac{{{\gamma _b}}}{{{\gamma _2}}}\frac{{{\delta \omega_1}{\Delta _2} - {\gamma _1}{\gamma _2}}}{{{g ^2}}}
\label{eq:14}
\end{equation}
In this expression the plus and minus before the first term correspond to two solutions. The solution with a minus is not stable for any parameter values, therefore, in what follows we consider only the solution with a plus.

Depending on the sign before the second term on the right side of Eq.~(\ref{eq:14}), either a soft excitation mode or a hard excitation mode is implemented in the laser. If the last term in the Eq.~(\ref{eq:14}) is negative, the nonzero solution can take place only when the expression~(\ref{eq:14}) becomes positive. The condition for the positivity of ${\left| {{a_2}} \right|^2}$ has the form
\begin{equation}
\Omega > {\Omega _{th}} = \frac{1}{g }\sqrt {\frac{{{\gamma _b}}}{{{\gamma _2}}}} \sqrt {{{\left( {{\delta \omega_1}{\gamma _2} + {\gamma _1}{\Delta _2}} \right)}^2} + {{\left( {{\gamma _1}{\gamma _2} - {\delta \omega _1}{\Delta _2}} \right)}^2}}
\label{eq:15}
\end{equation}
where we use the determination for ${\Omega _{ex}}$ (see the Eq.~(\ref{eq:13})). In this case, the condition that ${\left| {{a_2}} \right|^2}$ is zero coincides with the condition when the zero solution becomes unstable. Thus, when $\Omega  < {\Omega _{th}}$, the zero solution (${a_2} = b = 0$) is stable and the generation does not occur. When $\Omega  > {\Omega _{th}}$, the generation takes place and the intensity of the second optical mode, ${\left| {{a_2}} \right|^2}$, increases smoothly from zero [Figure~\ref{fig:1}a, b]. This behavior corresponds to the soft mode of excitation.

\begin{figure}[ht]
\centering
\includegraphics[width=0.5\linewidth]{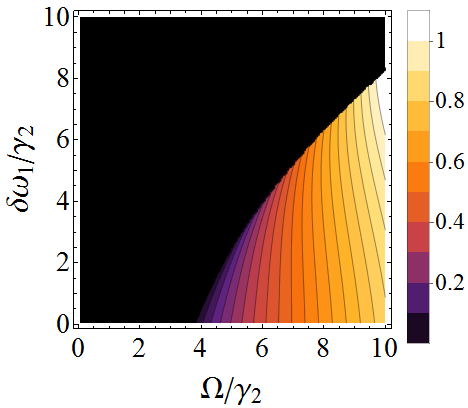}
\caption{The dependence of the intensity of second optical mode, ${\left| {{a_2}} \right|^2}$, on the amplitude external wave, $\Omega$, and $\delta \omega_1$ calculated using the Eqns.~(\ref{eq:5})-(\ref{eq:7}). Here $\delta \omega_2$ changes as $\delta \omega_1 + 2 \cdot 10^{-3}{\omega_0}$. All other parameters are the same as in Figure~\ref{fig:1}c.}
\label{fig:2}
\end{figure}

If the last term in the Eq.~(\ref{eq:14}) is positive, the nonzero solution appears when ${\Omega} = \Omega _{ex}$. However, this solution makes it stable only when ${\Omega}$ exceeds ${\Omega_{th}}$ (${\Omega_{th}} > {\Omega_{ex}}$). In this case, the intensity of the second optical mode, ${\left| {{a_2}} \right|^2}$, experiences a jump at the generation threshold [Figure~\ref{fig:1}c]. That is, the hard mode of excitation is observed in the laser.

\begin{figure*}[ht]
\centering
\includegraphics[width=\linewidth]{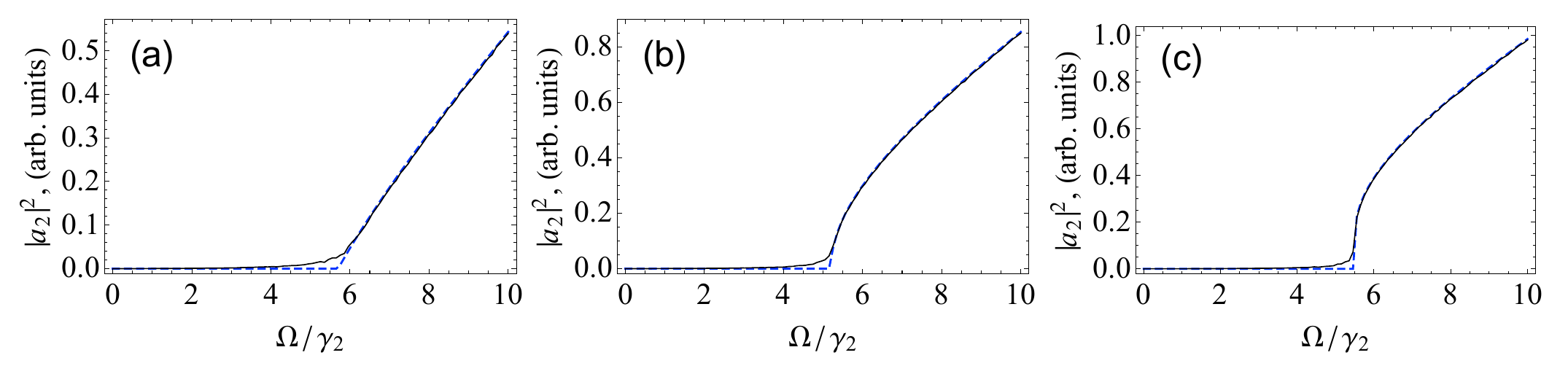}
\caption{The dependence of the intensity of second optical mode, ${\left| {{a_2}} \right|^2}$, on the amplitude of external wave, $\Omega$. The black line is calculated by using the equations with noise terms; the dashed blue line is calculated by using the Eqns.~(\ref{eq:2})-(\ref{eq:4}). The all parameters are the same as in Figures~\ref{fig:1}a, b, c, correspondingly.}
\label{fig:3}
\end{figure*}

The magnitude of the jump, $J$, is determined by the expression
\begin{equation}
\begin{array}{l}
J = \frac{1}{g }\sqrt {\frac{{{\gamma _b}}}{{{\gamma _2}}}} \sqrt {{{\Omega_{th} }^2} - \Omega _{ex}^2}  + \frac{{{\gamma _b}}}{{{\gamma _2}}}\frac{{{\delta \omega_1}{\Delta _2}-{\gamma _1}{\gamma _2}}}{{{g ^2}}} = \\
2{\gamma _b}\left( {\frac{{\delta {\omega _1}\left( {\delta {\omega _2} + {\omega _b}} \right)}}{{{\gamma _2} + {\gamma _b}}} - {\gamma _1}} \right)
\end{array}
\label{eq:16_0}
\end{equation}
where we use the Eq.~(\ref{eq:8_2}). This expression is positive when
\begin{equation}
{\delta \omega _1}\left( {\delta {\omega _2} + {\omega _b}} \right) > {\gamma _1}\left( {\gamma _2}+ {\gamma _b} \right)
\label{eq:16}
\end{equation}
It is important that $\delta \omega_{1,2}=\omega_{1,2} - \omega$ are frequency detunings and their magnitudes can be controlled by the frequency of the external wave, $\omega$. Thus, by changing the frequency of the external wave, we can move from the soft excitation mode to the hard excitation mode in the laser [Figure~\ref{fig:2}]. This makes it possible to achieve the jump-like change in the intensity of the second mode, ${\left| {{a_2}} \right|^2}$, at the lasing threshold.

From the Eq.~(\ref{eq:16_0}) it is seen that the magnitude of the jump is maximum when $\gamma_1$ and $\gamma_2$ tend to zero. The maximum jump value is given by

\begin{equation}
\max J = 2{\frac{\delta {\omega _1}\left( {\delta {\omega _2} + {\omega _b}} \right)}{g^2}}
\label{eq:20}
\end{equation}
If the condition of resonance between modes ($\omega_1 = \omega _2 + \omega_b$) is satisfied, then the expression (\ref{eq:20}) takes the form

\begin{equation}
\max J = 2{\frac{{\delta {\omega _1}}^2}{g^2}}
\label{eq:21}
\end{equation}

The abrupt change in the second mode intensity at the generation threshold can be used to create an extremely sensitive sensor operating on basis of the intercavity laser spectroscopy. Within this method, the changes in output-input curve caused by the addition of impurities in the cavity serves to determine the impurity concentration \cite{melentiev2017plasmonic,ma2014explosives,nechepurenko2018absorption}. The impurities can affect, for example, the frequencies and the relaxation rates of the modes. This leads to a change in the generation threshold and the shape of the laser curve \cite{melentiev2017plasmonic,ma2014explosives}. The sharper the impurity-induced change in the photon intensity, the higher the sensitivity of the sensor. In the considered case, the greatest change in the intensity takes place at the generation threshold ($\Omega  = {\Omega _{th}}$) when the condition~(\ref{eq:16}) is satisfied. Thus, the hard mode excitation in the laser increases sensitivity to changes in losses in the resonator, which can be used to detect low concentrations of absorbing impurities. That opens the way to create a highly sensitive laser operating based on the method of intracavity laser spectroscopy. In addition, the jump-like change in the second mode intensity near the generation threshold can be used to create optical transistor, in which a small change in the amplitude of the control signal, $\Omega$ (the external wave) leads to a large change in the detected signal (the radiation of the second mode).

It is known that noises greatly affects the laser curve near the generation threshold \cite{Scully}. To check that the hard mode excitation takes place in the laser even in the presence of noise, we simulate the Eqns.~(\ref{eq:2})-(\ref{eq:4}) with the additional noise terms \cite{Scully,Carmichael,Gardiner} (see \cite{mukhamedyanov2023subthreshold} for more details). We determine the noise amplitudes in accordance with the fluctuation-dissipation theorem \cite{Carmichael,Gardiner}. Our numerical simulation of the equations with the noises shows that even in the presence of noise, the abrupt change in [Figure~\ref{fig:3}c]. Thus, the noises does not lead to the disappearance of the hard excitation mode in the considered laser.

In conclusion, we have shown that the hard mode of excitation can take place in the system with optomechanical instability. This excitation mode is due to the existence of the additional phase condition that limits the range of parameters in which the nonzero solution exists. The phase condition can lead to the fact that the emerging nonzero solution immediately has an intensity greater than zero. In this case, the transition to the laser generation is accompanied by a jump-like increase in the amplitude of the second optical mode. In this regime, the laser can be used to create a highly sensitive sensors operating using the method of intracavity laser spectroscopy and the optical transistors.

\section*{Funding}
Russian Science Foundation (No. 20-72-10057).

\section*{Acknowledgments}
The study was financially supported by a Grant from Russian Science Foundation (project No. 20-72-10057).

\section*{Disclosures}
The authors declare no conflicts of interest.

\bibliography{sample}

\end{document}